\documentclass[aps, showkeys,nofootinbib,floatfix]{revtex4}

\usepackage{amssymb}
\usepackage{amsmath}
\usepackage{graphicx}
\usepackage{hyperref}
\usepackage{graphicx}

\begin{document}

\title{Solar system constraints on asymptotically flat IR modified Ho$\check{r}$ava gravity through light deflection}
\author{Molin Liu$^{1}$}
\email{mlliu@mail2.xytc.edu.cn}
\author{Junwang Lu$^{1}$}
\author{Benhai Yu$^{1}$}
\author{Jianbo Lu$^{2}$}
\email{lv$_$jb@163.com}
\affiliation{$^{1}$College of Physics and Electronic Engineering,
Xinyang Normal University, Xinyang, 464000, P. R. China\\
$^{2}$Department of Physics, Liaoning Normal University, Dalian, 116029, P. R. China}
\begin{abstract}
In this paper, we study the motion of photons around a Kehagias-Sfetsos (KS) black hole
and obtain constraints on IR
modified Ho$\check{r}$ava gravity without cosmological constant
($\sim \Lambda_{W}$). An analytic formula for the light deflection angle is obtained.
For a propagating photon, the deflection angle $\delta \varphi$ increases with
large values of the Ho$\check{r}$ava gravity
parameter $\omega$. Under the UV limit $\omega \longrightarrow
\infty$, deflection angle reduces to the result of usual
Schwarzschild case, $4GM/R$. It is also found
that with increasing scale of astronomical observation system the
Ho$\check{r}$ava-Lifshitz gravity should satisfy $|\omega M^2|>
1.1725 \times10^{-16}$ with $12\%$ precision for Earth system,
$|\omega M^2| > 8.27649 \times 10^{-17}$ with $17\%$ precision for
Jupiter system and $|\omega M^2| > 8.27650\times 10^{-15}$ with
$0.17\%$ precision for solar system.
\end{abstract}



\maketitle

\section{Introduction}
It is known that the Lagrangian of original Lifshitz scalar field theory is
\begin{equation}\label{lifshitz}
\mathcal{L} = \int d^2 x d t \left[\left(\partial_t \phi\right)^2 - \lambda_{L} \left(\Delta^2 \phi\right)^2\right],
\end{equation}
which has a line of fixed points parameterized by $\lambda_{L}$ with
anisotropic scale invariance \cite{Lifshitz}. So such fixed points
are the so-called Lifshitz points. Recently, Ho$\check{r}$ava
adopted one Lifshitz scale in time and space, and obtained a
renormalizable gravity theory at these Lifshitz point
\cite{Horava1,Horava2,Horava3}. It is known also as
Ho$\check{r}$ava-Lifshitz (HL) gravity. It exhibits a broken Lorentz
symmetry at the short distances, but it reduces to usual General
Relativity (GR) at the large distances due to these non-contribution
of higher derivative terms with particular $\lambda = 1$. The
parameter $\lambda$ controls the contribution of the extrinsic
curvature trace. However, it is possible that HL at large distance
could deviate from GR for the incomplete diffeomorphism invariance.
In this paper, we only consider the simple case of $\lambda = 1$.
Since HL gravity theory was put forth, HL gravity is intensively
investigated in many aspects. They are developments of basic
formalism
\cite{Sotitiou,Visser,CCai,Chenen,Orlando,CCCai,Nishioka1122,LiLIli,Charmousis,Sotiriou111,calagni222,Blas333,Iengo,Germani11,Mukohyama1,Appignani1,Kluson1,Afshordi,Kobakhidze1,Kobakhidze2},
cosmology \cite{Kiritsis, Mukohyama, Brandenberger, Piao,
Mukohyama2, Kalyana2, chen1, Wang1, Nojiri, Cai1, Wang2,
Kobayashi1}, dark energy and dark matter
\cite{Saridakis,Park,Mukohyama11}, various black hole solutions and
their thermodynamics
\cite{Saridakis,Park,Mukohyama11,Nastase,Cai22,Myung,Cai33,Mann,Bottacant,Castillo,Peng111,Colgain,Lee,Kim111,Park22,Lu,Kehagias,Koutsoumbas1,Koutsoumbas2,Majhi222}
and so on.

With the subsequent developments of the HL gravity, the following problem raised in front of us: how does the matter fields influence the geometry? Several researchers made efforts to find the analogue of the matter energy-momentum tensor acting gravitational source as GR. The study of
its geodesic in HL theory is the first step for finding a solution. Hence, some pioneering works addressed this aspect with various methods. They are the optical limit of a scalar field
theory \cite{Capasso}, super Hamiltonian formalism \cite{Rama},
foliation preserving diffeomorphisms \cite{Mosaffa},
Lorentz-violating Modified Dispersion Relations \cite{Sindoni} as
well as observational kinematic constraints below mentioned \cite{chenjuhua,Iorio1,Iorio2,Harko11}. In particular, in the framework of the optical limit Capasso and Polychronakos \cite{Capasso} gave us a comprehensive study of particles in the HL gravity. Based on
the optical limit, the deformed geodesic equation was derived from a
generalized Klein-Gordon action. Such results clearly show that the fact that the particles move along geodesic, which depend on the mass of particles, is not granted. Deviations from geodesic motion appear both in flat and Schwarzschild-like spacetimes.
This type deformed kinematics could allow the presence of superluminal particle for particular $\lambda$ ( or $\tilde{\lambda}$'s in Ref.\cite{Capasso}) which could not give any restriction on the velocity. The particles with given $\tilde{\lambda}$'s could be always accelerated to any velocity under an action of constant force. Else, the higher is the energy of the particle, the larger will be the deviation from
the geodesic motion determined by the low-energy four dimensional
metric $g_{\mu\nu}$. Similar results deviations from GR also are
found in \cite{Rama,Mosaffa,Sindoni} with various above-mentioned
approaches.

On the other hand, a specific case of asymptotically flat IR modified HL gravity under the cosmological constant
limit $\Lambda_{W} \longrightarrow 0$ in HL gravity is proposed by
Kehagias and Sfetsos \cite{Kehagias}. It reduces to GR gravity in IR
regime according to coupling $\lambda$.The massless excitation with
two propagating polarizations emerges in the Minkowski vacuum, which
is admitted naturally, with the transverse traceless part of the
perturbation. At the most important, one static spherically
symmetric black hole solution of HL gravity i.e. KS black hole is
obtained, which can be treated as counterpart of the Schwarzschild
case in GR. Comparing its geometry with GR, it is also found that
this solution has the same Newtonian and post-Newtonian limits which
ensures its soundness when it is confronted with the astronomical
observational data. Since then, there are some works involving the
phenomenology of the deformed HL black hole, such as strong field
gravitational lensing \cite{chensongbai1, Konoplya}, black hole's
quasi-normal modes \cite{Konoplya, chensongbai2}, timelike Geodesic
Motion \cite{chenjuhua}, thin accretion disk \cite{Harko1}. In
Konoplya's work \cite{Konoplya}, the gravitational lensing and
quasi-normal modes are considered, which are potentially observable
properties of black holes beyond the post-Newtonian corrections. It
is found that the bending angle is smaller and the real oscillation
frequency of quasinormal modes (QNM) is larger than that of GR case.
The QNM can be lived longer in this deformed HL black hole. In
Chen's work \cite{chensongbai1}, the angular position and
magnification of the relativistic images are obtained with strong
field gravitational lensing. Unlike the former Konoplya's work
\cite{Konoplya}, a scheme to distinguish a deformed HL black hole
from a Reissner-Norstr$\ddot{o}$m black hole is introduced.
Comparing with the latter work, the angular position decreases more
slowly with bigger HL gravity parameter $\omega$. Else, angular
separation increases more rapidly. In Chen's another work
\cite{chensongbai2}, it is also found that the scalar perturbations
decay more slowly in the HL gravity. In Ref. \cite{chenjuhua}, the
time-like geodesic motion is analyzed with the effective potential
in this deformed HL gravity. In Harko's work \cite{Harko1}, by using
the accretion disk properties, one can obtain some important
properties such as the energy flux, the temperature distribution,
the emission spectrum and so on. Unlike the usual GR, there are some
particular signatures appeared in the electromagnetic spectrum which
maybe supply a observable path for HL gravity.

With the development of KS black hole, there are some works referring to
field parameters constraints such as the solar system orbital
motions \cite{Iorio1}, the extrasolar planets and double pulsar
\cite{Iorio2},  the solar system three classical tests \cite{Harko11}
and so on. In \cite{Iorio1,Iorio2} lower bounds on $\omega$ are obtained via the post-Newtonian motion equation.

 As we know that, the asymptotically flat HL gravity with $\lambda =1$
reduces to the Schwarzschild space with the limit of
$\omega\longrightarrow+\infty$. So in the UV limit, the field
parameter $\omega$ should be large enough to ensure the UV
Ho$\check{r}$ava gravity is in conformity with GR gravity well.
Motivated by above various factors, one aim of this paper is to find
the lower limit of HL gravity field parameter $\omega$. Considering
the last work \cite{Harko11}, the numerical constraints results are
obtained directly from integrating metric function by using the
static and spherically symmetric metric. Hence, another aim of this
paper is to find the analytic expression of light deflection. We
adopt one approach of Lagrange analysis and study the light
deflection in the KS black hole in HL gravity under the weak field
and the slow motion approximation. Then, we also compare our results
with the observational data (long baseline radio interferometry,
Jupiter measurement, Hipparcos satellite) and we also obtain corresponding
constraints on HL field parameter $\omega$ for solar system, Jovian
planet system and Earth system, respectively.

This paper is organized as follows. In section II, we present the
action of HL gravity and spherically symmetric Kehagias and Sfetsos
black hole. In section III, we calculate the general solution to
motion equation of photons in deformed HL gravity. In section IV, we
use astronomical observations to constrain the HL gravity.
In Section V we compare our results with other constraints obtained from inside and outside solar system. Section VI is devoted to the conclusions. We adopt the signature $(-, +, +, +)$ and put $\hbar$, $c$, and $G$
equal to unity.
\section{The asymptotically flat black hole solution in the deformed HL gravity}
In this section, we review briefly the KS black hole solutions under
the limit of $\Lambda_{W} \longrightarrow 0$ with running constant
$\lambda = 1$ in the IR critical point $z = 1$. The spacetime geometry is, in the ADM formalism \cite{ADD},
\begin{equation}\label{ADMmetric}
    d s^2 = - N^2 d t^2 + g_{ij} \left( d  x^i + N^i d t\right)\left(d x^j + N^j dt\right),
\end{equation}
where the classical scaling dimensions of the field are
\begin{equation}\label{scalingdimensions}
    [g_{i j}]_{s} = 0, \ \ [N]_{s} = 0, \ \ [N^{i}]_{s} = 2.
\end{equation}
The action for the fields of HL theory is

\begin{eqnarray}\label{action1}
\nonumber S &=& \int dt d^3 x \sqrt{g} N \bigg{\{} \frac{2}{\kappa^2} \left(K_{ij}K^{ij} - \lambda K^2 \right) - \frac{\kappa^2}{2 w^4} C_{ij} C^{ij} + \frac{\kappa^2 \mu}{2 w^2} \epsilon^{ijk} R_{il}^{(3)} \nabla_{j} R^{(3)l}_{\ \ \ \ k}\\
    && \ \ \ \ \ \ \ \ \ \ \ \ \ \ \ \ \ \ \ -\frac{\kappa^2 \mu^2}{8} R_{ij}^{(3)}R^{(3)ij} + \frac{\kappa^2 \mu^2}{8 \left(1 - 3\lambda\right)} \left(\frac{1 - 4 \lambda}{4}\left(R^{(3)}\right)^2 + \Lambda_{W} R^{(3)} - 3 \Lambda^2_{W}\right) + \mu^4 R^{(3)}\bigg{\}},
\end{eqnarray}
where the second fundamental form, extrinsic curvature $K_{ij}$, and
the Cotton tensor $C^{ij}$ are given as follows,
\begin{eqnarray}
  K_{ij} &=& \frac{1}{2N} \left(\frac{\partial }{\partial t}g_{ij} - \nabla_i N_j - \nabla_j N^i\right), \label{extrinsiccurvature}\\
  C^{ij} &=& \epsilon^{ikl} \nabla_k \left(R^{(3)j}_{l} - \frac{1}{4} R^{(3)} \delta_{l}^{j}\right). \label{Cottontensor}
\end{eqnarray}
Here, $\kappa$, $\lambda$, $w$ are dimensionless coupling constants and $\mu$, $\Lambda_{W}$ are dimensional parameters having dimensions of masses $[\mu] = 1$, $[\Lambda_{w}] = 2$. The last term of metric (\ref{action1}) represents a soft violation of the detailed balance condition. In the limit of $\Lambda_{W} \longrightarrow 0$, we can obtain a deformed action as follows,
\begin{eqnarray}
    S &=& \int dt d^3 x \left(\mathcal{L}_0 + \mathcal{L}_1\right),\label{deformedaction1}\\
    \mathcal{L}_0 &=& \sqrt{g} N \bigg{\{}\frac{2}{\kappa^2} \left(K_{ij}K^{ij} - \lambda K^2\right)\bigg{\}},\label{deformedaction2}\\
    \mathcal{L}_1 &=& \sqrt{g} N \bigg{\{} \frac{\kappa^2 \mu^2 \left(1 - 4\lambda\right)}{32\left(1 - 3\lambda\right)} \mathcal{R}^2 - \frac{\kappa^2}{2 w^4}\left(C_{ij} -\frac{\mu w^2}{2} R_{ij}\right) \left(C^{ij}-\frac{\mu w^2}{2} R^{ij}\right) + \mu^4 \mathcal{R} \bigg{\}}.\label{deformedaction3}
\end{eqnarray}
Comparing these with the usual GR ADM formalism, we can obtain the speed of light $c$, the Newton's constant $G$ and the cosmological constant $\Lambda$
\begin{equation}\label{cGlambda}
    c = \frac{\kappa^2 \mu}{4} \sqrt{\frac{\Lambda_{W}}{1 - 3\lambda}},\ \ G = \frac{\kappa^2}{32\pi c},\ \ \Lambda = \frac{3}{2} \Lambda_{W}.
\end{equation}
For the particular case of $\lambda = 1$ with $\omega = 16 \mu^2/\kappa^2$, a spherically symmetric black hole solution is presented by Kehagias and Sfetsos \cite{Kehagias}, which also is corresponding to an asymptotically flat space,
\begin{equation}\label{Kehagiassolution}
    d s^2 = -f(r) d t^2 + \frac{d r^2}{f(r)} + r^2 \left(d \theta^2 + \sin^2 \theta d \varphi^2 \right).
\end{equation}
The lapse function is
\begin{equation}\label{metricfunction}
    f(r) = 1 + \omega r^2 - \sqrt{r\left(\omega^2 r^3 +4 \omega M\right)},
\end{equation}
where the parameter $M$ is an integration constant with dimension $[M] = -1$. Using the null hypersurface condition, one can find there are two horizons, inner $r_-$ and outer event horizon $r_+$ in this space,
\begin{equation}\label{horizons}
    r_{\pm} = M \left(1 \pm \sqrt{1 - \frac{1}{2 \omega M^2}}\right).
\end{equation}
The Hawking temperature is
\begin{equation}\label{hawkingtemperature}
    T_{H} = \frac{\sqrt{2 M^2 \omega - 1}}{\pi \sqrt{2\omega} (r_+^2 + \frac{1}{\omega} +\sqrt{r_+^4 + 4 \frac{M r_+}{\omega}})}.
\end{equation}
The corresponding black hole thermodynamics and its generalized uncertainty principle are studied in Ref. \cite{Myungadda}. It is also found that if $r \gg (M/\omega)^{1/3}$, this KS solutions will reduce to usual Schwarzschild case and the lapse function can be rewritten as
\begin{equation}\label{metricfunctionssss}
    f \approx 1 - \frac{2M}{r} + \mathcal{O}(r^{-4}).
\end{equation}
Apparently, the former integration constant $M$ is the real mass of black hole.

\section{motion of photons in HL gravity: general solution}
\begin{figure}
  \includegraphics[width=3.5 in]{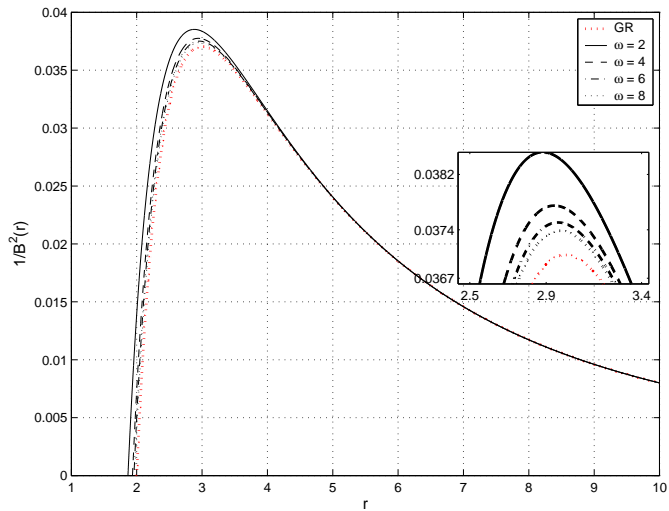}\\
  \caption{The effective potentials $1/B^2 (r)$ of photons with unit mass and various parameters $\omega$. The usual Schwarzschild case in GR is denoted with bigger dotted red line.}\label{potential}
\end{figure}

By using the lapse function of KS black hole (\ref{Kehagiassolution}), the usual Lagrangian of test particle is given as
\begin{equation}\label{lagrangian}
    2 \mathcal{L}_p = - f(r) \dot{t}^2 + \frac{\dot{r}^2}{f(r)} + r^2 \left(\dot{\theta}^2 +\sin^2 \theta \dot{\varphi}^2\right),
\end{equation}
where the overdot represents differentiation with respect to the affine parameter $\xi$ along the geodesics. The Euler-Lagrange equations below give us the kinetic equation of photons,
\begin{equation}\label{ELequations}
    \frac{d}{d \xi} \left(\frac{\partial \mathcal{L}_p}{\partial \dot{x}^{\mu}}\right) - \frac{\partial \mathcal{L}_p}{\partial x^{\mu}} = 0.
\end{equation}
Without loss of generality, it is assumed that the observer is confined to orbits with $\theta = \pi/2$ and $\dot{\theta} = 0$. Hence, the Lagrangian $\mathcal{L}_p$ reduce to
\begin{equation}\label{lagrangian0}
    2 \mathcal{L}_p = - f(r) \dot{t}^2 + \frac{\dot{r}^2}{f(r)} + r^2 \dot{\varphi}^2.
\end{equation}
According to the Euler-Lagrange equations (\ref{ELequations}) and reduced Lagrangian $\mathcal{L}_p$ (\ref{lagrangian0}), we can identify two constant motions for cyclic coordinates $\varphi$ and $t$,
\begin{eqnarray}
  f (r) \dot{t} &=& E, \label{cycliccoor1} \\
  r^2 \dot{\varphi} &=& L, \label{cycliccoor2}
\end{eqnarray}
where $E$ and $L$ are the linear momentum and angular momentum at infinite distance. According to the condition for photons $g_{\mu \nu } p^{\mu} p^{\nu} = 0$ where $p^{\mu} \equiv \frac{d x^{\mu}}{d \xi} = \dot{x}^{\mu}$ is the photons momentum, we can obtain a relation of parameters ($\dot{t}, \dot{r}, \dot{\varphi}$) as,
\begin{equation}\label{trverphi}
    -f(r) \dot{t}^2 + \frac{1}{f(r)} \dot{r}^2 + r^2 \dot{\varphi}^2 = 0.
\end{equation}
Dividing Eq. (\ref{trverphi}) by $\dot{\varphi}^2$, we can obtain motion equation as
\begin{equation}\label{motioneeqq}
    - f(r) \left(\frac{\partial t}{\partial \varphi}\right)^2 + \frac{1}{f(r)}\left(\frac{\partial r}{\partial \varphi}\right)^2 + r^2 = 0.
\end{equation}
Substituting  Eqs.(\ref{cycliccoor1}), (\ref{cycliccoor2}) into  Eq. (\ref{motioneeqq}), we can obtain the final photons' orbital equation
\begin{equation}\label{orbitalequation}
    \left(\frac{1}{r^2} \frac{d r}{d \varphi}\right)^2 = \frac{1}{b^2} - \frac{1}{B^2(r)},
\end{equation}
where $E$ and $L$ are coupled via $b= L/E$ which represents the effective sighting range or impact parameter. The corresponding photon effective potential $1/B^2(r)$ is given as
\begin{equation}\label{photonpotential}
    B^2 (r) = \frac{r^2}{1 + \omega r^2 - \sqrt{r\left(\omega^2 r^3 +4 \omega M\right)}}.
\end{equation}
Fig.(\ref{potential}) clearly illustrates that the potential becomes smaller for a larger $\omega$. One also could find that with the bigger HL gravity parameter $\omega$ the curves tends toward the usual GR Schwarzschild result which is
drew with bigger red dotted line in Fig.(\ref{potential}). This phenomenon is in agreement with the Schwarzschild limit $r^3 \gg M/\omega$ \cite{Kehagias}, in which for the fixed variables $r$ and the central mass parameter $M$, the rewritten Schwarzschild limit is $\omega \gg M/r^3$. Hence, the bigger $\omega$ will lead to
a final Schwarzschild case.

The constant impact parameter $b$ in Eq. (\ref{orbitalequation}) can be eliminated via calculating the derivation with respect to $\varphi$. So the photon
Binet equation can be obtained
\begin{equation}\label{traqu2}
    \frac{d^2 u}{d \varphi^2} = -u + 3 u^2 \sqrt{\omega M^2} \left(\omega M^2 + 4 u^3\right)^{-\frac{1}{2}},
\end{equation}
where $u = M/r$. Considering the fact of that all astronomical observations can be explained very well by the Schwarzschild gravity in solar system, we adopt the Schwarzschild approximation $r \gg (M/\omega)^{1/3}$. Hence, the final general solution to photon orbital equation (\ref{orbitalequation}) is written directly as
\begin{equation}\label{generalsolution}
    u = u_0\cos\varphi + u_0^2 \left(1 + \sin^2 \varphi\right) +\frac{a}{2}\left(\varphi \sin\varphi -\frac{1}{8} \cos 3 \varphi -\frac{1}{120} \cos 5\varphi\right).
\end{equation}
Here, the parameters $u_0$ and $a$ are $u_0 = G M /R$ and $a = -15 u_0^5/4 \omega M^2$
where $R$ is the solar radius and $M$ is the solar mass.
In the right hand side of Eq. (\ref{generalsolution}), the first term $u_0\cos\varphi$ is a
straight line normal to pole axis without considering the influence
of gravity, the second term $u_0^2 \left(1 + \sin^2 \varphi\right)$
is the standard form coming from general relativity gravity, and the
last one $\frac{a}{2}\left(\varphi \sin\varphi -\frac{1}{8} \cos 3
\varphi -\frac{1}{120} \cos 5\varphi\right)$ is an improvement of HL
Gravity. Part of some related calculations are given in the following section.
\section{Solution to photon orbital equation with Ho$\breve{r}$ava-Lifshitz Gravity}
Under Schwarzschild approximation $r \gg (M/\omega)^{1/3}$, the photon Binet equation Eq.(\ref{traqu2}) reduces to a solvable form,
\begin{equation}\label{appendix001}
    \frac{d^2 u}{d \varphi^2} = -u +3u^2 -6 \frac{u^5}{\omega M^2}.
\end{equation}
Considering the small $u$, the high-order small quantities $u^2$ and $u^5$ can be ignored firstly. So we can obtain the zeroth-order approximate solution
\begin{equation}\label{apppendixzerothorder}
    u = u_0 \cos \varphi,
\end{equation}
where parameter $u_0 = GM/R$ is a constant about solar system. Mathematically, this solution is a straight line normal to the pole axis. In order to simplify the calculation, we only replace the high-order terms with the zeroth-order solution in Eq. (\ref{appendix001}) not the fussy iteration step by step. Hence, one can get
\begin{equation}\label{apppendix002}
    \frac{d^2 u}{d \varphi^2}= -u + 3u_0^2 \cos^2 \varphi - \frac{6 u_0^5}{\omega M^2} \cos^5 \varphi.
\end{equation}
After decreasing the powers of $\cos^5 \varphi$ term, Eq.(\ref{apppendix002}) can be reduced a new form
\begin{equation}\label{appendix003}
    \frac{d^2 u}{d \varphi^2} + u = 3 u_0^2 \cos^2 \varphi - \frac{3 u_0^5}{8\omega M^2} \cos 5\varphi -\frac{15u_0^5}{8\omega M^2} \cos 3\varphi -\frac{15u_0^5}{4\omega M^2} \cos \varphi.
\end{equation}
Because the right hand of Eq.(\ref{appendix003}) includes four terms, we can decompose this ordinary differential equation (\ref{appendix003}) into corresponding four sub-equations to solve. Here, we list these particular solutions in the \textbf{Table} \ref{table1}.
\begin{table*}[!h]
\caption{The particular solutions to Eq.(\ref{appendix003}).}
\label{table1}
\begin{center}
\begin{tabular}{|c|c|}
     \hline
 $\text{Sub-equation}$&$\text{Particular solution}$\\
     \hline
    $\frac{d^2 u}{d \varphi^2} + u = 3 u_0^2 \cos^2 \varphi$ & $u = u_0^2 \left(1 + \sin^2 \varphi\right)$\\
     \hline
    $\frac{d^2 u}{d \varphi^2} + u = - \frac{3 u_0^5}{8\omega M^2} \cos 5\varphi$& $u = -\frac{a}{240} \cos 5\varphi$\\
     \hline
    $\frac{d^2 u}{d \varphi^2} + u = -\frac{15u_0^5}{8\omega M^2} \cos 3\varphi$ & $u = -\frac{a}{16} \cos 3\varphi$\\
     \hline
    $\frac{d^2 u}{d \varphi^2} + u = -\frac{15u_0^5}{4\omega M^2} \cos \varphi$ & $u = \frac{a}{2} \varphi \sin\varphi$\\
\hline
\end{tabular}
\end{center}
\end{table*}

Hence, the total particular solution is
\begin{equation}\label{totalparti}
    u^{*} = u_0^2 \left(1 + \sin^2 \varphi\right) +\frac{a}{2}\left(\varphi \sin\varphi -\frac{1}{8} \cos 3 \varphi -\frac{1}{120} \cos 5\varphi\right).
\end{equation}
And the first-order approximate solution to Eq.(\ref{appendix001}) is given as
\begin{equation}\label{approxi006}
    u = u_0\cos\varphi + u_0^2 \left(1 + \sin^2 \varphi\right) +\frac{a}{2}\left(\varphi \sin\varphi -\frac{1}{8} \cos 3 \varphi -\frac{1}{120} \cos 5\varphi\right).
\end{equation}

\begin{figure}
  \includegraphics[width=3.4 in]{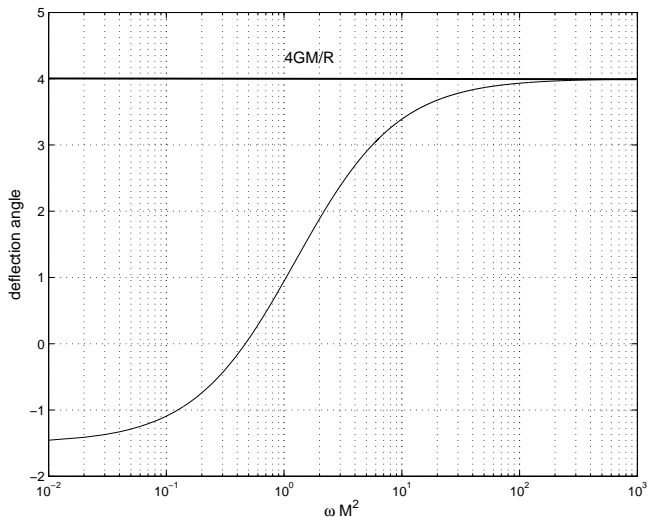}\\
  \caption{Deflection angle $\delta\varphi$ versus HL Gravity parameter $\omega M^2$ with $u_0 = 1$. The solid bold line denoted by $4GM/R$ is the usual Schwarzschild case [57].}\label{deflectionangleeps}
\end{figure}

The azimuth angles of zeroth-order approximate solutions (\ref{apppendixzerothorder})
are $\pm\pi/2$ at very great distance. However, the azimuth
angles of the first-order ones (\ref{approxi006}) are $\pm(\pi/2 +
\beta)$ in $u = 0$. The deflection angle $\beta$ is a small quantity
which satisfies
\begin{equation}\label{app120}
 -u_0 \sin\beta + u_0^2 (1 + \cos^2\beta) + \frac{a}{2} \left[\left(\frac{\pi}{2} + \beta\right)\cos \beta -\frac{1}{8} \sin 3\beta +\frac{1}{120} \sin 5\beta\right] = 0.
\end{equation}
We use the Taylor expansions of sine and cosine functions and keep
the basic term. The deflection angle is written formally as
\begin{equation}\label{app110}
   \delta\varphi =2\beta = \frac{4 G M}{R} - \left(\frac{5 G u_0^4}{R M} + \frac{15 \pi G u_0^3}{8 R M}\right) \cdot \frac{1}{\omega} + \left(\frac{25 G u_0^8}{4 R M^3} + \frac{75 \pi G u_0^7}{32 R M^3}\right) \cdot \frac{1}{\omega^2} + \mathcal{O}\left(\frac{1}{\omega^3}\right),
\end{equation}
where the first term $4GM/R$ is the usual result of the pure Schwarzschild spacetime \cite{book}. The other high-order correction items containing parameter $\omega$ are the contribution of HL Gravity. It affects the light deflection only through the particular solution (\ref{totalparti}), but the zeroth order approximate (\ref{apppendixzerothorder}) is unchanged. Furthermore, the well
asymptotically behavior of deflection angle, $\delta \varphi
\longrightarrow 4 G M/R$ under the limit $\omega \longrightarrow
+\infty$, could ensure UV HL gravity does not break the classical GR
gravity. The deflection angles versus $\omega M^2$ are illustrated
in Fig.\ref{deflectionangleeps} in which $\delta\varphi$ increases
with bigger $\omega M^2$. Furthermore, the upper limit of this curve
is the usual Schwarzschild case i.e. $4 GM/R$.

\section{Constraints on Ho$\breve{r}$ava-Lifshitz Gravity from astronomical observations}
In this section, we will use the observational data to confine the Ho$\breve{r}$ava-Lifshitz Gravity based on the deflection angle Eq.(\ref{app110}). It should be noticed that in order to analyse exactly the results Eq.(\ref{app110}), we move from the natural units used so far to the SI system by restoring $c$ and $G$ throughout the relevant formulas. In order to obtain experimental constraints from the light deflection
result, the final light deflection angle $\delta \varphi$ should be
expressed in terms of the deviation $\Delta_{LD}$ from the general
relativity prediction $\delta \varphi _{GR}$ for the Sun.
\begin{equation}\label{deviation}
\delta \varphi = \delta \varphi_{GR}(1+\Delta_{LD}),
\end{equation}
where $\delta \varphi_{GR} = 4 G M /c^2 R$. The best available
constraints on $\delta \varphi _{GR}$ come from long-baseline radio
interferometry which shows that $|\Delta_{LD}| \leq 0.0017$
by Robertson et al. \cite{Robertson} and Lebach et al. \cite{Lebach}. Submitting the deflection angles Eq.(\ref{app110}) into
Eq.(\ref{deviation}) and neglecting the higher-order terms, we can therefore infer an lower limit of Ho$\breve{r}$ava-Lifshitz gravity parameter
\begin{equation}\label{lowerlimit}
|\omega M^2|>8.27650\times10^{-15},
\end{equation}
for the solar system. However, it is important to bear in minds that
the parameter $\omega$ characterizing the HL gravity metric
(\ref{Kehagiassolution}) is not a naturally universal constant like
$G$ or $c$, but it may in principle vary from one space to another. HL
gravity as an alternative to 4D general relativity is therefore best
constrained by the application of two or more tests to the same
system.

With this in mind, according to the deviation $|\Delta_{LD}| \leq 0.17$ from the Jupiter measurement of light deflection by Treuhaft and Lowe \cite{Treuhaft}, we can obtain
\begin{equation}\label{Jupiter}
    |\omega M^2| > 8.27649\times 10^{-17}.
\end{equation}
It has also been proposed by Gould \cite{Gould} to measure the light deflection due to the Earth by using the Hipparcos satellite, with an estimated
precision of $12\%$, which leads a sensitive values for the Earth system
\begin{equation}\label{12342121}
|\omega M^2|> 1.1725 \times10^{-16}.
\end{equation}

\section{comparison with others constrains from solar system and extrasolar planets}
Here we compare our results with the constraints obtained by other researchers using solar system and exoplanetary data.  The solar system tests include perihelion
precessions \cite{Iorio1,Harko11}, radar echo delays and light bending \cite{Harko11}. The   extrasolar planets have been recently used by Iorio and Ruggiero \cite{Iorio2} as well.

In the former classical solar system tests \cite{Harko11}, the constrains are obtained via the analysis of arbitrary spherically symmetric spacetimes \cite{Harko11}, which is different from the dynamic Lagrangian analysing in this paper.
It should be noted
that the parameters $\omega_0$ and $\psi_0$ used in \cite{Iorio1,Iorio2,Harko11} coincide with our $\omega M^2$. For the perihelion precession case, the circular orbit $u_0$
is determined by the nonlinear algebraic equation (51) in Ref. \cite{Harko11}
which could bring us numerical solutions. So using
the planet Mercury data, one can obtain the corresponding result
$\omega M^2 \sim 7 \times 10^{-16}$ with the 43 arcsec per century of
precession angle immediately, which is in the same order with our result
of Hipparcos-Earth system, i.e. $1.2 \times 10^{-16}$. For the radar echo delays case, the time interval between emission and return receiver is determined by the integration of lapse function,
\begin{equation}\label{conclusionadd}
    \Delta t = \int_{-l_1}^{l_2} \bigg{\{}e^{\left[\lambda (\sqrt{y^2 + R^2}) - \nu (\sqrt{y^2 + R^2})\right]/2} - 1\bigg{\}}.
\end{equation}
Though this general expression is simple, it is hard to solve and obtain a explicit form of time delays, except for the numerical solution, since there is a square root term in the KS lapse function. The numerical integration shows the result $\omega M^2 \sim 4 \times 10^{-15}$ which has an exact identical order of magnitude with our long-baseline radio interferometry result, i.e. $8.27650 \times 10^{-15}$ under analytic analysis perspective. For the light deflection case, it is slightly emphasized that, in spite of considering light bending both in Ref.\cite{Harko11} and this paper, the based methods have nothing in common. Furthermore, the analysis of light bending based on the clear analytic expression of deflection angle (\ref{app110}) is grasped in this paper. Then, we have a look at the relating results in Ref. \cite{Harko11}. The photon's motion equation satisfies the integral expression (55) of Ref. \cite{Harko11}. Hence, the deflection angle is $\Delta \phi = 2 |\phi(x_0) - \phi(\infty)| - \pi$. Like the former radar echo delays case, since there are some square root terms, the numerical solution is obtained rather than analytic one. Submitting the classical observational value $\Delta \phi = 1.7275$ arcsec, the result shows parameter $\omega_0 = 10^{-15}$, which is the same order as our results $\omega M^2 \sim 8.3 \times 10^{-15}$ based on the analytic perspective. Meanwhile, this results of the constraints on asymptotically flat IR modified Ho$\check{r}$ava gravity are compatible with the constraints in Jupiter and Earth systems. The corresponding compatibilities of parameter $\omega M^2$ are in the orders of $1 \times 10^{-15} (\text{Harko\ and\ Kovacs}) > 8.27650 \times 10^{-15} (\text{Solar\ \ system}) > 1.1725 \times 10^{-16} (\text{Earth\ \ system}) > 8.27649 \times 10^{-17} (\text{Jupiter\ \ system})$. Obviously, the constraints on  Ho$\check{r}$ava gravity field parameter $\omega$ are more restrictive via the analytic method of this paper.

Then we consider the orbital motions tests \cite{Iorio1,Iorio2} in the following words. It should be noticed that in the approximation of weak field and
slow motion, orbital motion mainly refers to the secular precession of the longitude for the pericenter test particle which has not perihelion at all. So orbital motion is different from that of the general relativistic Einstein's pericentre precession. The constraints of orbital motion are divided into two parts, inside \cite{Iorio1} and outside \cite{Iorio2} our solar system. The secular precessions
of the longitude of the pericentre $\varpi$ are calculated for a test particle by using the Gauss and Lagrange perturbative approaches. Then, based on the EPM2008 ephemerides
constructed from a data set of 550000 observations ranging from 1913 to
2008 \cite{Pitjeva}, they compared their results with
the corrections $\Delta \dot{\varpi}$ to the standard Newtonian/
Einsteinian planetary perihelion precessions and found the lower bounds on KS solutions. For the case of solar system planets \cite{Iorio1}, inferior planets (Mercury, Venus, Earth, Mars) and outer planets (Jupiter, Saturn, Uranus, Neptune, Pluto), the lower limit of $\omega M^2$ is
\begin{equation}\label{conclusionadd}
    |\psi_0| \geq \bigg| - \frac{3 (GM)^{7/2} (4 + e^2)}{2 \left[\Delta \dot{\varpi} \pm \delta (\Delta \dot{\varpi})\right]_{Max} c^6 a^{9/2} (1 - e^2)^3}\bigg|,
\end{equation}
where $a$ is the semi-major axis and $e$ is the planet's
eccentricity. So the corresponding constraints on HL gravity ranging
from $|\psi_0| \geq 10^{-12} $ to $|\psi_0| \geq 10^{-24} $ are
shown in following tables \cite{Iorio1}. Interestingly, the larger
value $8 \times 10^{-12}$ in their work \cite{Iorio1} coming from
Mercury is about 1000 times more than that of our result
Eq.(\ref{lowerlimit}) coming from radio interferometry. For the same
Earth, we also find that Iorio and Ruggiero's result $6 \times
10^{-14}$ is almost 513 times more than that of our result
Eq.(\ref{12342121}). In contrast, our result Eq.(\ref{Jupiter}) is
almost 30 times more than Iorio and Ruggiero's result $3 \times
10^{-18}$ for the same Jupiter plant.
\begin{table}[!h]\label{11}
\caption{Lower bounds on $|\psi_0|$ of of the inner planets and outer planets in solar system \cite{Iorio1}.}
\begin{center}
\begin{tabular}{|c|c|c|c|c|c|}
     \hline
           inner & Mercury & Venus & Earth & Mars & --- \\ \cline{2-6}
           planets & $8 \times 10^{-12}$ & $6 \times 10^{-14}$ & $6 \times 10^{-14}$ & $8 \times 10^{-15}$ & ---\\
        \hline
        outer & Jupiter & Saturn & Uranus & Neptune & Pluto\\ \cline{2-6}
        planets & $3 \times 10^{-18}$ & $1 \times 10^{-18}$ & $2 \times 10^{-22}$ & $2\times 10^{-23}$ & $8 \times 10^{-24}$\\
     \hline
\end{tabular}
\end{center}
\end{table}

Then we consider the other orbital motions type constraints coming from the extrasolar planets shown also by Iorio and Ruggiero \cite{Iorio2}. Like their early solar system work \cite{Iorio1}, the same standard Gauss perturbative approach was adopted. The correction to the Keplerian period is
\begin{equation}\label{addcorrKep}
    P_{\omega_0} = \frac{4\pi (GM)^{5/2} (1 + \frac{5}{4} e^2)}{\omega_0 c^6 a^{3/2} (1 - e^2)^{5/2}},
\end{equation}
where $\omega_0 = \omega M^2 = \psi_0$. By using the limit of $e
\longrightarrow 0$, one can equates the centripetal acceleration to
the Newton + KS gravitational acceleration for a circular orbit. The
final expression of $\omega_0$ in HL gravity is
\begin{equation}\label{addextrosolar}
    \omega_0 = \frac{4\pi (GM)^{5/2}}{\Delta P c^6 d^{3/2}},
\end{equation}
where $d$ represents the current fixed radius of the circular orbit and $\Delta P$ is the different value of correction orbital period to the third Kepler law. Comparing these
with the phenomenologically determined orbital periods of the transiting extrasolar
planet HD209458b ``Osiris", the result $\omega_0 \geq 1.4 \times 10^{-18}$ which is about one fiftieth of our minimum Jupiter planet value Eq.(\ref{Jupiter}). Obviously, for the extrasolar planets give us a more rigorous constraints.

\section{conclusion}
In this paper, we have studied the constraints on HL gravity from
light deflection astronomical observations including long-baseline
radio interferometry \cite{Robertson, Lebach}, Jupiter measurement
\cite{Treuhaft}, Hipparcos satellite \cite{Gould}. Here we summarize our findings.

We need briefly specify the parameters $\lambda$ and $\Lambda_W$
firstly. $\lambda$ is dimensionless coupling constant which controls
the contribution of the extrinsic curvature trace. In this paper, we
only consider the specific case of $\lambda = 1$ which is
corresponding to usual GR case at large distances. The parameter
$\Lambda_{W}$ is the cosmological constant of HL gravity. For the
various HL gravity models with Lifshitz points, $\Lambda_{W}$ has
the different reduced values. For example, for $Z = 1$ HL gravity in
this paper, its value is $\Lambda_{W} = 3\Lambda/2$ where $\Lambda$
is the cosmological constant of usual 4D GR. For $Z = 4$ HL gravity,
its values are $\Lambda_{W} = 3\Lambda/2$ for (3 + 1) dimensions and
$\Lambda_{W} = \Lambda$ for (4 + 1) dimensions \cite{CCai}.
Considering the asymptotically flat IR modified HL gravity, we adopt
$\Lambda_{W} = 0$ in this paper.

By using the conventional Lagrange approach for the calculus of variations,
the the motion of photons is studied in the deformed HL gravity.
The photons' orbital equation (\ref{orbitalequation}) is coupled
with the key HL Gravity parameter $\omega$ in the potential function
$1/B^2(r)$. Hence, a scattering problem in a central force field occurs here.
It is found that the photon collisional potential, which is larger for smaller $\omega$
is larger than in the usual GR case. Furthermore, for large values of $\omega$, the HL gravity  reduces to the Schwarzschild case which is also justified by the Kehagias and
Sfetsos's Schwarzschild limit \cite{Kehagias}.

All the classical tests in the solar system are in agreement with GR, especially those by long-baseline radio interferometry \cite{Robertson,Lebach}, Jupiter measurement \cite{Treuhaft}, Hipparcos satellite \cite{Gould}. So in this paper we use the
Schwarzschild limit \cite{Kehagias} and  the difficult Eq.(\ref{traqu2}) is directly solved according to the linear
superposition principle of ordinary differential equation. In the final result of Eq. (\ref{app110}), we find that the deflection angle value increases monotonically for large values of the
HL gravity parameter $\omega M^2$ (or $\omega$).  For a small magnitude of $\omega$ (about $10^{-17}\sim 10^{-15}$ here), the modification of HL gravity is dominated by the second term ($\sim 1/\omega$) according to Eq.(\ref{app110}). So the result of HL gravity is less than the usual GR value $4GM/R $. This relation of HL and GR is also clearly illustrated in FIG.\ref{deflectionangleeps}. The similar behavior will be
found in the strong field gravitational lensing
\cite{chensongbai1,Konoplya}. Finally, the constraints on HL gravity
from long-baseline radio interferometry \cite{Robertson, Lebach},
Jupiter measurement \cite{Treuhaft}, Hipparcos satellite
\cite{Gould} are $ |\omega M^2|>8.27650\times10^{-15}$, $ |\omega
M^2| > 8.27649\times 10^{-17}$ and $|\omega M^2|> 1.1725
\times10^{-16}$ for Solar system, Jupiter system and Earth system,
respectively.

For the comprehensive consideration of the constraints on
asymptotically flat IR modified HL gravity, we also compare our
results with others similar works including usual GR tests
\cite{Harko11}, orbital motions inside \cite{Iorio1} and outsider
\cite{Iorio2} solar system. Firstly, for the comparison with usual
GR tests \cite{Harko11}, we find our constraint from Hipparcos-Earth
system ($\sim 10^{-16}$) is concordant with the result of perihelion
precession of Ref.\cite{Harko11}. Meanwhile, our constraint from
long-baseline interferometry ($\sim 10^{-15}$) is in agreement with
their result of radar echo delays case. However, for the same light
deflection test, our results could give a more lower bound than
their value ($\sim 10^{-15}$). Secondly, for the comparison with
orbital motions inside solar system \cite{Iorio1}, we find that our
result ($\sim 10^{-16}$) gives a lower bound, about two orders of
magnitude, than that of their result ($\sim 10^{-14}$) for the same
Earth system. However, our result ($\sim 10^{-15}$) gives a higher
bound, about three orders of magnitude, than that of their result
($\sim 10^{-18}$) for the same Jupiter system. Thirdly, for the
comparison with orbital motions outsider solar system \cite{Iorio2},
the lowest bound from our Jupiter case ($\sim 10^{-17}$) is higher
one orders than that of HD209458b ``Osiris" ($\sim 10^{-18}$). In
general, according to the above situation, the HL gravity field parameter $\omega_0$ or $\psi_0$ should be
in the range of $[10^{-12}, 10^{-24}]$. The constraint from solar system mainly concentrates in the
range of $[10^{-17}, 10^{-15}]$, except for the orbital
motions. On the contrary, the constraint
from orbital motions distributes very separately, in which the
largest bound ($\sim 10^{-12}$) and the lowest bound ($\sim
10^{-24}$) come from in Mercury system and Pluto system,
respectively.

\end{document}